\documentclass[prl,aps,twocolumn,amsmath,amsmath,amssymb,
    superscriptaddress,longbibliography]{revtex4-1}

\usepackage{graphicx}
\usepackage[dvipsnames]{xcolor}
\usepackage[colorlinks,linkcolor=blue,citecolor=blue,urlcolor=blue]{hyperref}
\usepackage{stix}
\usepackage[utf8]{inputenc} 
\usepackage[T1]{fontenc}

\renewcommand{\vec}[1]{\mathbf{#1}}
\newcommand{\gvec}[1]{\boldsymbol{#1}}
\newcommand{\en}{\varepsilon}

\begin{document}

\title{Berry Curvature Signatures in Chiroptical Excitonic Transitions}

\author{Samuel Beaulieu}
\email{samuel.beaulieu@u-bordeaux.fr}
\affiliation{Universit\'e de Bordeaux - CNRS - CEA, CELIA, UMR5107, F33405 Talence, France}

\author{Shuo Dong}
\email{dong@fhi-berlin.mpg.de}
\affiliation{Beijing National Laboratory for Condensed Matter Physics, Institute of Physics, Chinese Academy of Sciences, Beijing 100190, China}
\affiliation{Fritz-Haber-Institut der Max-Planck-Gesellschaft, Faradayweg 4-6, 14195 Berlin, Germany}

\author{Viktor Christiansson}
\affiliation{Department of Physics, University of Fribourg, 1700 Fribourg, Switzerland}

\author{Philipp Werner}
\affiliation{Department of Physics, University of Fribourg, 1700 Fribourg, Switzerland}

\author{Tommaso Pincelli}
\affiliation{Fritz-Haber-Institut der Max-Planck-Gesellschaft, Faradayweg 4-6, 14195 Berlin, Germany}
\affiliation{Institut für Optik und Atomare Physik, Technische Universität Berlin, Strasse des 17 Juni 135, 10623 Berlin, Germany}

\author{Jonas D. Ziegler}
\affiliation{Institute of Applied Physics and Würzburg-Dresden Cluster of Excellence ct.qmat, Technische Universität Dresden, 01062 Dresden, Germany}

\author{Takashi Taniguchi}
\affiliation{Research Center for Materials Nanoarchitectonics, National Institute for Materials Science,  1-1 Namiki, Tsukuba 305-0044, Japan}

\author{Kenji Watanabe}
\affiliation{Research Center for Electronic and Optical Materials, National Institute for Materials Science, 1-1 Namiki, Tsukuba 305-0044, Japan}

\author{Alexey Chernikov}
\affiliation{Institute of Applied Physics and Würzburg-Dresden Cluster of Excellence ct.qmat, Technische Universität Dresden, 01062 Dresden, Germany}

\author{Martin Wolf}
\affiliation{Fritz-Haber-Institut der Max-Planck-Gesellschaft, Faradayweg 4-6, 14195 Berlin, Germany}

\author{Laurenz Rettig}
\affiliation{Fritz-Haber-Institut der Max-Planck-Gesellschaft, Faradayweg 4-6, 14195 Berlin, Germany}

\author{Ralph Ernstorfer}
\affiliation{Fritz-Haber-Institut der Max-Planck-Gesellschaft, Faradayweg 4-6, 14195 Berlin, Germany}
\affiliation{Institut für Optik und Atomare Physik, Technische Universität Berlin, Strasse des 17 Juni 135, 10623 Berlin, Germany}

\author{Michael Schüler}
\email{michael.schueler@psi.ch}
\affiliation{Laboratory for Materials Simulations, Paul Scherrer Institut, 5232 Villigen PSI, Switzerland}
\affiliation{Department of Physics, University of Fribourg, 1700 Fribourg, Switzerland}

\begin{abstract}

The topology of the electronic band structure of solids can be described by its Berry curvature distribution across the Brillouin zone. We theoretically introduce and experimentally demonstrate a general methodology based on the measurement of energy- and momentum-resolved optical transition rates, allowing to reveal signatures of Berry curvature texture in reciprocal space. By performing time- and angle-resolved photoemission spectroscopy of atomically thin WSe$_2$ using polarization-modulated excitations, we demonstrate that excitons become an asset in extracting the quantum geometrical properties of solids. We also investigate the resilience of our measurement protocol against ultrafast scattering processes following direct chiroptical transitions.

\end{abstract}
\date{\today}
\maketitle

\section{Introduction}

Electron transport and dynamics in periodically-ordered solids are governed by intrinsic quantum mechanical properties, such as the electronic band structure and the interaction between electrons, phonons, and other quasiparticles. The quantum geometrical properties of the Bloch wavefunction, manifesting as Berry curvature (property that reflects handedness of Bloch electrons), band topology, Fermi-liquid transport properties~\cite{karplus_hall_1954}, current-noise characteristics~\cite{neupert_measuring_2013}, or the geometric origin of superfluidity in flat-band systems~\cite{julku_geometric_2016}, play a fundamental role in all of these microscopic mechanisms. More generally, the quantum geometry of Bloch electrons is of capital importance, as it provides key insights into the intricate interplay between quantum mechanics and materials' electronic properties. Recently, the link between the quantum geometry and light-matter interaction has entered the stage, providing insights into physical mechanisms underlying peculiar optoelectronic responses of topological materials~\cite{Yao08, ahn_riemannian_2022,topp_light-matter_2021-1,morimoto_topological_2016-1,puente-uriona_ab_2023}. 

However, a momentum-resolved measurement of Bloch electrons' quantum geometry still remains a grand challenge. A direct approach -- exploiting the close link between the quantum geometry and light-matter interaction -- has been introduced in the context of cold atoms, where paradigmatic model systems can directly be implemented.  Indeed, since interband transition dipole matrix elements are equivalent to the Berry connection~\cite{resta_quantum-mechanical_1998}, the rate of transitions from occupied to unoccupied bands upon resonant monochromatic irradiation has been shown to be a direct measure of the underlying quantum geometry~\cite{tran_probing_2017}. In particular, the circular dichroism in the absorption is a fingerprint of a Chern insulating state, which has been demonstrated out-of-equilibrium~\cite{schuler_tracing_2017}, for fractional quantum Hall systems~\cite{repellin_detecting_2019}, and in optical lattices~\cite{asteria_measuring_2019}. However, applying this approach to diagnose a material's quantum geometry is not straightforward. Indeed, for systems with locally nonzero but globally vanishing Berry curvature, the total (momentum-integrated) optical oscillator strength does not provide any specific information on the quantum geometrical properties. However, it has been predicted that quantum geometric information can in principle be extracted from dichroism in the \emph{momentum-resolved} optical oscillator strength~\cite{Yao08,von_gersdorff_measurement_2021,chen_measurement_2022}. Even if the connection between k-resolved optical oscillator strengths of interband transitions and Berry curvature is already established~\cite{Yao08}, an associated experimental measurement protocol for extracting local quantum geometric information of materials is still missing. Tackling this problem requires going beyond standard optical spectroscopic probes, as they lack momentum resolution.  

This is where angle-resolved photoemission spectroscopy (ARPES)~\cite{damascelli_probing_2004,lv_angle-resolved_2019} has entered the stage. Indeed, signatures of local Berry curvature in solids can be extracted by using circularly polarized ionizing radiation~\cite{Wang11, Razzoli17, Schuler20, Cho18,cho_studying_2021,unzelmann_momentum-space_2021}. The basic principle of this approach is based on the close relation between Berry curvature and orbital angular momentum (OAM). Intuitively, it has been shown that OAM is linked with a self-rotation of the initial state, which is reflected in the dipole selection rules in the ARPES matrix elements -- circular dichroism can emerge because of propensity rules in photoemission for electrons co- or counter-rotating with circularly polarized light. However, despite its feasibility, extracting information on the Berry curvature from photoemission transition dipole matrix elements is not straightforward in practice due to the influence of the experimental geometry~\cite{Schonhense90} and effects of complicated photoelectron final states~\cite{bentmann_strong_2017,moser_toy_2023}.

\begin{figure}[t]
\centering
\includegraphics[width=\columnwidth]{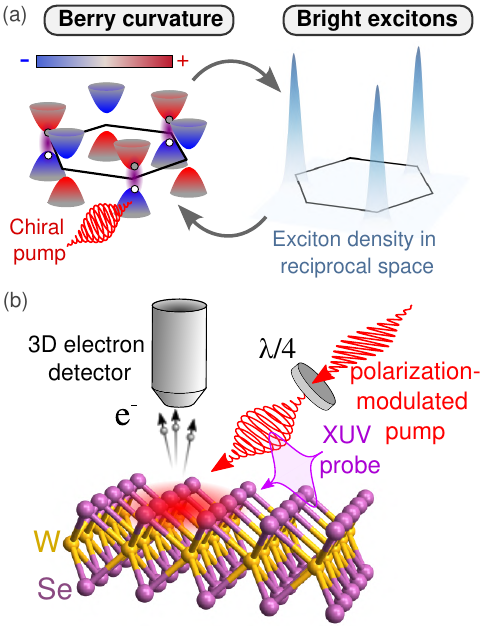}
\caption{\textbf{Illustration of Berry curvature texture and exciton population along with the schematic of the experimental measurement protocol}. (a) Single particle top valence band and bottom conduction band of WSe$_2$ close to the K valley. Due to the Berry curvature (represented by color shading), the electrons and holes created upon photoexcitation possess intrinsic orbital angular momentum. The optical transition rate is modulated by the chirality of excitons and of the pump pulse and serves as a probe of the Berry curvature. (b) Sketch of the experimental setup, featuring a polarization-modulated IR pump and linearly-polarized XUV probe pulses. Photoelectrons are collected by a time-of-flight momentum microscope detector. }
\label{fig:intro}
\end{figure}

The extension to the time domain using time-resolved ARPES (trARPES) -- a powerful technique to measure out-of-equilibrium band structures and excited states of crystalline solids -- allows, in principle, to directly measure the momentum-resolved optical interband transition rate. For example, momentum-resolved linear dichroism in bilayer MoS$_2$ in trARPES has been shown to reveal intralayer single-particle hopping~\cite{Volckaert19}. Extending this approach to chiral (circular) excitations allows to translate the cold-atom concept of the dichroism of the depletion rate into pump dichroism of the population of the unoccupied states.
However, the rich ultrafast dynamics within the photoexcited material, leading to the redistribution of optically prepared excited states both in energy and momentum, can blur the direct relationship between measured photoemission intensities and momentum-resolved optical oscillator strength. In particular, electron-electron and electron-phonon scattering can smear out the initial energy-momentum distribution of pump-induced excited states on the femtosecond timescale. In addition, many-body excitations such as excitons or correlated in-gap states are often the dominating excitation channel, which, at first sight, seems to obscure the link between optical transition rates and quantum geometry. 

In this work, we in turn use the many-body excitations to extract of quantum geometrical properties of solids. In particular, by exploiting the optical selection rules for chiral valley-excitons, we map out the Berry curvature texture of the prototypical atomically thin transition metal dichalcogenide (TMDC) WSe$_2$. We show that the measurement of the momentum-resolved chiroptical oscillator strength, using optical pump polarization-modulation in trARPES, allows us to access the electronic wavefunction's quantum geometry texture in materials.

\begin{figure*}[t]
\begin{center}
\includegraphics[width=\textwidth]{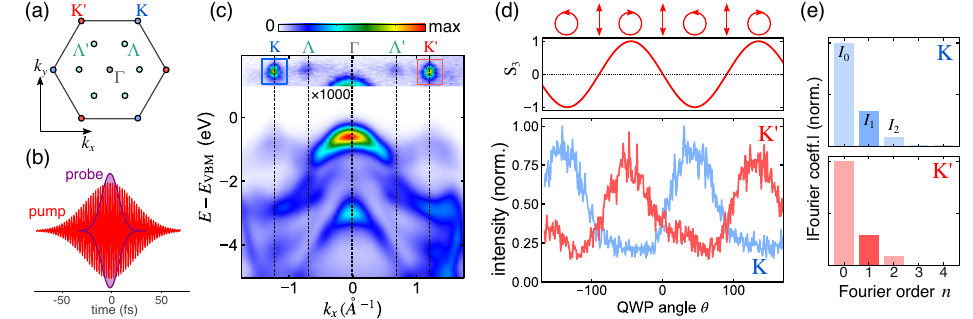}
\caption{\textbf{Optical polarization-modulated pump-probe photoemission in monolayer WSe$_2$}. (a) Sketch of the Brillouin zone of WSe$_2$ with the high-symmetry points. (b) Sketch of the overlapping pump and probe pulses. (c) Optical polarization-averaged trARPES signal along $k_x$ (K-$\mathrm{\Lambda}$-$\Gamma$-$\mathrm{\Lambda}$'-K'). The intensity has been multiplied by 1000 for unoccupied states. (d) Ellipticity factor (Stokes parameter $S_3$) of the pump pulse, which is controlled by the continuous rotation of quarter-wave-plate (QWP) angle $\theta$ (top panel), along with the ellipticity-resolved photoemission intensity of excited states around the K and K' points. (e) The absolute value of the Fourier coefficients associated with the polarization-modulated photoemission intensities from excitonic states in (d). The highlighted coefficient ($n=1$) is associated with the helicity-swap frequency, i.e. captures the effects of circular polarization.}
\label{fig:exp_bands}
\end{center}
\end{figure*}

\section{Results}

Monolayer WSe$_2$ (ML-WSe$_2$) possesses broken inversion symmetry and strong spin-orbit coupling, leading to locked spin, orbital, and valley degrees of freedom~\cite{Zhang14}. These symmetry considerations imply peculiar valley-selective optical selection rules, leading to strong circular dichroism~\cite{mak2012control, zeng2012valley, Cao2012} -- a property that is at the heart of our approach. These material systems are also characterized by specific orbital angular momentum and Berry curvature texture in reciprocal space. In addition, monolayer WSe$_2$ has a direct band gap at the two inequivalent K and K' valleys.
Due to the reduced screening resulting from its atomically thin nature, its excitons have large binding energies and dominate their optical responses, even at room temperature. As a result, strongly bound (hundreds of meV) bright excitons comprised of electrons and holes in the vicinity of K/K' in the top valence and bottom conduction band are formed (known as A-excitons) upon resonant photoexcitation. These strongly bound excitons are stable against momentum scattering for relatively long time scales. In contrast, typical band-to-band single-particle excitations at higher energy are subject to electron-electron and electron-phonon scattering on the femtosecond time scale. The key concept of our approach is summarized in Fig.~\ref{fig:intro}(a): the Berry curvature of the valence and conduction bands is tied to OAM. Therefore, excitons as bound states of electrons and holes become chiral excitations, whose population is determined by whether the chirality of the pump aligns with their intrinsic chirality. In turn, the exciton population (as measured from trARPES) is characteristic of the Berry curvature of the underlying valence and conduction band.
While the chirality of excitons has been discussed in terms of winding numbers~\cite{cao_unifying_2018} and from first principles~\cite{caruso_chirality_2022}, its use for the reconstruction of the Berry curvature texture is an unexplored territory. 

\subsection{Experiments}
In our trARPES setup, bright K/K' excitons are resonantly prepared at room temperature by a resonant near-infrared (NIR) pump pulse (760~nm, $\hbar\omega_\mathrm{IR}=1.63$~eV, $\sim$ 45~fs full width at half maximum (FWHM) duration). Electrons with momenta corresponding to first Brillouin zone (Fig.~\ref{fig:exp_bands}(a)) are ejected from the sample (ML-WSe$_2$ on thin hBN flake on a slightly Nb-doped rutile TiO$_2$ (100) substrate -- for more details, see Methods) through the photoelectric effect induced by linearly p-polarized XUV pulses (57~nm, $\hbar\omega_\mathrm{pr}=21.7$~eV and $\sim 20$~fs FWHM duration).  Measurements are performed at the pump-probe overlap ($\Delta t =0$) to maximize the signal emerging from bright excitons (Fig.~\ref{fig:exp_bands}(b)), while simultaneously minimizing the contribution of ultrafast scattering processes following photoexcitation. We recorded two-color (NIR+XUV) ARPES spectra while continuously rotating the quarter-wave plate (QWP) angle $\theta$, leading to a pump polarization-modulation from left-hand circularly polarized (LCP) to linearly s-polarized to right-hand circularly polarized (RCP) (top  panel in Fig.~\ref{fig:exp_bands}(d)). This continuous polarization-modulated photoemission measurement protocol is analogous to a lock-in detection scheme. Indeed, using Fourier analysis, this measurement scheme allows us to isolate signals which are modulated at the helicity-swap frequency, efficiently rejecting all other frequency components coming from e.g. linear dichroism, experimental geometry, or artifacts (imperfection of the waveplate, misalignments, etc.). The photoemission data are acquired using a time-of-flight momentum microscope, which allows to detect each photoelectron as a single event, as a function of NIR quarter-waveplate angle ($\theta$), resulting in 4D photoemission intensity data -- $I(k_x,k_y,E,\theta)$.  More information about the experimental setup can be found in the methods section.

A typical ARPES signal along K-$\Gamma$-K' high symmetry direction (pump polarization-integrated) is shown in Fig.~\ref{fig:exp_bands}(c). Bright excitons directly manifest themselves in Fig.~\ref{fig:exp_bands}(c) as strongly localized (in energy-momentum space) pump-induced signals ($E-E_{\mathrm{VBM}} \sim \hbar\omega_\mathrm{IR}$) at the Brillouin zone (BZ) boundaries in the trARPES spectra~\cite{dong_direct_2021, madeo_directly_2020}. In addition, photoemission intensity at $\Gamma$, which can be attributed to laser-assisted photoemission (LAPE)~\cite{Glover96}, as well as signatures of momentum-indirect dark excitons at the $\Sigma$ valleys are also visible in Fig.~\ref{fig:exp_bands}(c). 

In Fig.~\ref{fig:exp_bands}(d), we show the modulation of the photoemission signal from bright excitons at K (K'), momentum and energy-integrated for the three equivalent valleys, as a function of the NIR quarter-wave plate angle. Note that before summing the signal emerging from the three equivalent K and K' valleys, we made sure that the modulation in each equivalent valley was following the same trend. Signals originating from excitons located around both K and K' valleys are strongly modulated, with a dominating oscillation component with a 180$^{\circ}$ period (helicity-swap period). The $\pi$-phase shift between the modulations of the K and K' excitons indicates that these quasiparticles are created upon the absorption of light with opposite chirality, RCP and LCP, respectively. The $\frac{\pi}{2}$-phase with an identical population of K and K' valley excitons reflects equal excitation with a linearly polarized pump. These results already indicate that the phase of the exciton population modulation encodes some information related to their intrinsic valley pseudospin degree of freedom. 

From the full $\theta$-dependent intensity, we can perform a Fourier analysis of the experimentally measures signals in Fig.~\ref{fig:exp_bands}(d). Besides a non-oscillating background (encoded in the $n=0$ component), the $n=1$ Fourier component is dominant at both K and K' (Fig.~\ref{fig:exp_bands}(e)), consistent with the modulation of the light chirality. The $n=2$ Fourier coefficient is originating mainly from linear dichroism, i.e. the modulation between s- and p- components of pump pulses. 
Because we recorded four-dimensional ARPES data $I(k_x,k_y,E,\theta)$, we have access to the polarization-modulated ($\theta$) photoemission signal for each energy ($E$) and momenta ($k_x,k_y$) coordinates. We can thus perform the energy- and momentum-resolved Fourier analysis, i.e. compute the Fourier components for each voxel
\begin{align}
    \label{eq:fourier_intensity}
    I_n(k_x,k_y,E) = \frac{1}{2\pi}\int^\pi_{-\pi} d\theta\, e^{-2i n \theta}I(k_x,k_y,E,\theta)  \ .
\end{align}

This procedure yields complex quantities containing the full information on the excitation with linearly polarized photons (encoded in $I_2(k_x,k_y,E)$), and circular dichroism (encoded in $I_1(k_x,k_y,E)$). $I_0(k_x,k_y,E)$ and the imaginary part of $I_1(k_x,k_y,E)$ computed from the experimental data are shown in Fig.~\ref{fig:berry_fourier}(c)-(d), respectively. 

While the dominant components of $\mathrm{Im}[I_1(k_x,k_y,E)]$ are strong signals at BZ corners with alternating signs between K and K' valleys, suggesting qualitatively some similarity with the OAM and Berry curvature texture, the detailed understanding of the origin of these features requires some theoretical analysis, which is done in the following sections.

\subsection{Theory of exciton signatures}
We treat excitons in the electron-hole basis, expanding the many-body state 
\begin{align}
    \label{eq:exciton_def}
    |\Psi^\mathrm{exc}_{\vec{p}\lambda}\rangle = \sum_{\vec{k}\alpha\beta} Y^\lambda_{\alpha\beta}(\vec{p},\vec{k}) c^\dagger_{\vec{k}+\vec{p}\alpha} c_{\vec{k}\beta} | \Psi_0\rangle \ .
\end{align} 
Here, $\vec{p}$ denotes the center-of-mass momentum of the exciton (different states labeled by $\lambda$), while $c^\dagger_{\vec{k}+\vec{p}\alpha}$ ($c_{\vec{k}\beta}$) creates an electron (a hole) in the conduction (valence) band $\alpha$ ($\beta$) with corresponding momentum; $|\Psi_0\rangle$ is the ground state. The envelope function $Y_{\alpha\beta}(\vec{p},\vec{k})$ -- its Fourier transform limited size can be experimentally measured~\cite{Dong21,Man21,Schmitt2022} -- describes the localization of the excitons. For excitons in TMDCs, $Y_{\alpha\beta}(\vec{p},\vec{k})$ is  strongly localized around $\vec{k}=$K/K' for bright excitons, while for the dark excitons, $\vec{k}$ is localized around K/K' ($\Lambda$) for holes (electrons). 

In the linear-response regime, the population $P_\mathrm{exc}$ of the bright excitons is obtained from Fermi's Golden rule (assuming atomic units)
\begin{align}
    \label{eq:exciton_fermi}
    P^\lambda_\mathrm{exc}(\theta) = S^2(\omega_\mathrm{IR} - E^\lambda_\mathrm{exc}) \left|\vec{e}_\mathrm{IR}(\theta) \cdot \vec{M}^\lambda \right|^2 \ ,
\end{align}
where $E^\lambda_\mathrm{exc}$ is the energy of the two A-excitons relative to the ground state, while $\vec{e}_\mathrm{IR}(\theta)$ denotes the polarization of the NIR pump pulse. The dipole matrix element of the excitons is given by $\vec{M}^\lambda$, while $S(\omega)$ stands for the Fourier transform of the envelope of the pump pulse (all other constant prefactors have been absorbed into $S(\omega)$). 
Combining the wave-function~\eqref{eq:exciton_def} and the exciton population~\eqref{eq:exciton_fermi} with the trARPES formalism~\cite{freericks_theoretical_2009,schuler_theory_2021} and assuming that the exciton population stays constant over the duration of the probe pulse, one finds
\begin{align}
    \label{eq:trARPES_theory}
    I_{\vec{p}\lambda}(k_x,k_y,E,\theta) &\propto 
    g(\en_\beta(\vec{k} - \vec{p}) + E^\lambda_\mathrm{exc}(\vec{p}) + \omega_\mathrm{pr} - E)     
    \nonumber \\ &\quad \times 
    P^\lambda_\mathrm{exc}(\vec{p},\theta)\sum_{\beta} \left|Y^\lambda_{\alpha\beta}(\vec{p},\vec{k})\right|^2  \ .
\end{align}
Here, $\en_\beta(\vec{k})$ denotes the energy of the valence bands,  $\omega_\mathrm{pr}$ the photon energy of the probe pulse, 
and $E$ the energy of the final states, all entering a Gaussian function $g(\omega)$ whose width is determined by the duration of the probe pulse. We also include the dark excitons ($\vec{p}\ne 0$) in Eq.~\eqref{eq:trARPES_theory}, as they get populated on a sub-100~fs time scale due to electron-phonon scattering~\cite{Bertoni16}. Neglecting photoemission matrix elements, the experimental intensity $I(k_x,k_y,E,\theta)$ is obtained from Eq.~\eqref{eq:trARPES_theory} by summing over all exciton momenta $\vec{p}$ in the first BZ. 

Apart from enabling a direct comparison with the experimental results, our theory allows us to trace the dependence of the trARPES intensity on the QWP angle $\theta$ back to the exciton population. For the bright excitons, the Fourier components~\eqref{eq:fourier_intensity} are thus determined by $I_n(k_x,k_y,E) \propto \int^\pi_{-\pi} d\theta / (2\pi)\, e^{-2i n \theta} |\vec{e}_\mathrm{IR}(\theta) \cdot \vec{M}^\lambda |^2$. Working out the pump polarization $\vec{e}_\mathrm{IR}(\theta)$ in the given experimental geometry, the $n=1$ Fourier component is given by
\begin{align}
    \label{eq:fourier_n1}
    \mathrm{Im}\left[I_{n=1}(k_x,k_y,E)\right] & \propto \frac{\cos\alpha}{2} \mathrm{Im}\left[(M^\lambda_x)^* M^\lambda_y \right] \ ,
\end{align}
where $\alpha$ denotes the angle of incidence.
The combination of matrix elements in Eq.~\eqref{eq:fourier_n1} is directly proportional to the circular dichroism:
\begin{align}
\label{eq:fourier_n1_cd}
    \mathrm{Im}\left[I_{n=1}(k_x,k_y,E)\right] 
    \propto -\frac{\cos\alpha}{4}\left(P^\mathrm{LCP}_\mathrm{exc} - P^\mathrm{RCP}_\mathrm{exc}
    \right) \ .
\end{align}
Here, $P^\mathrm{LCP}_\mathrm{exc}$ ($P^\mathrm{RCP}_\mathrm{exc}$) is the exciton population that would be generated by a pump with LCP (RCP) polarization in \emph{normal} incidence. The component $I_{n=2}$ is related to linear dichroism. 
In summary, sweeping over the QWP angle $\theta$ and Fourier transforming the ARPES signal provides direct access to energy- and momentum-resolved chiroptical (pump) circular dichroism in normal incidence, while the experimental geometry enters only as a prefactor. 

\subsection{Impact of Berry curvature on excitons}
\begin{figure*}[t]
\begin{center}
\includegraphics[width=\textwidth]{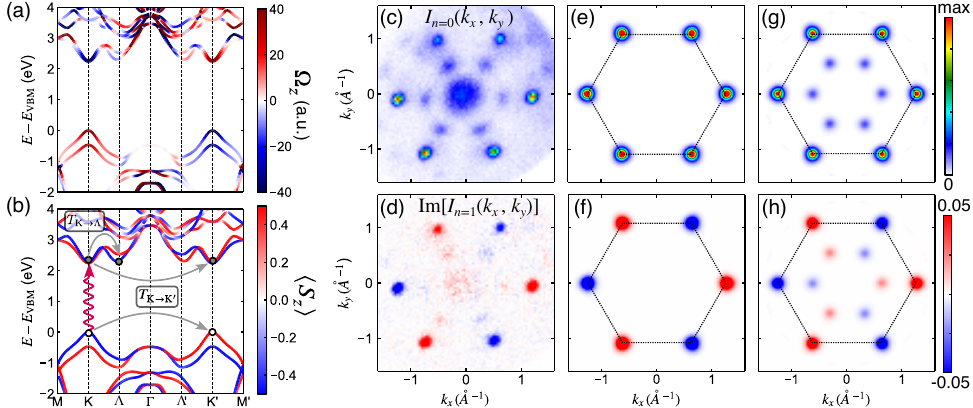}
\caption{\textbf{Berry curvature, spin texture, and Fourier components of the dichroic signal of excitons.} (a) Berry curvature of monolayer WSe$_2$ along indicated high-symmetry points. (b) Spin expectation value texture $\langle S_z\rangle$ along the same high-symmetry points. The arrows illustrate the pump excitation and the relevant exciton scattering processes in the electron and hole picture. (c) Polarization-averaged photoemission intensity (equivalent to the $n=0$ Fourier component), energy-integrated over the excited state's region. (d) Imaginary part of the $n=1$ Fourier component $I_{n=1}(k_x,k_y,E)$ (energy-integrated as in (c)). (e), (f) Theoretical predictions (without any scattering) of the $n=0$ and $n=1$ Fourier components of the intensity corresponding to (c), (d). (g), (h) Theoretical predictions where the inter-valley scattering has been included.}
\label{fig:berry_fourier}
\end{center}
\end{figure*}

To trace the impact of the quantum geometry on the pump-induced exciton population, we analyze the dipole transition matrix element $\vec{M}^\lambda$ of the bright excitons in Eq.~\eqref{eq:exciton_fermi}. The light-matter coupling is expressed through the coupling of the pump electric field $\vec{E}_\mathrm{p}(t)$ and the polarization operator $\hat{\vec{P}}$: $\hat{H}_{lm} = - \vec{E}(t) \cdot \hat{\vec{P}}$. For interband transitions, the matrix elements of $\hat{\vec{P}}$ in the basis of Bloch states $|\psi_{\vec{k}\alpha}\rangle$ are given by $\vec{A}_{\alpha\alpha^\prime}(\vec{k}) = \langle \psi_{\vec{k}\alpha} | \vec{r} | \psi_{\vec{k}\alpha^\prime}\rangle$. With the modern theory of polarization~\cite{resta_quantum-mechanical_1998} we can identify the matrix elements $\vec{A}_{\alpha\alpha^\prime}(\vec{k})$ with the Berry connections $i \langle u_{\vec{k}\alpha} | \nabla_{\vec{k}} u_{\vec{k}\alpha^\prime} \rangle$ ($|u_{\vec{k}\alpha}\rangle$ is the cell-periodic part of the Bloch wave-function). Combining this with the exciton wave-function~\eqref{eq:exciton_def}, the exciton transition matrix element becomes
$\vec{M}^\lambda = \sum_{\vec{k}\alpha\beta} Y^\lambda_{\alpha\beta}(\vec{k}) \vec{A}_{\alpha\beta}(\vec{k})$. Inserting into Fermi's Golden rule~\eqref{eq:exciton_fermi} and exploiting the localization in momentum space, we obtain the leading contribution to the circular dichroism $P^\mathrm{CD}_\mathrm{exc} =  P^\mathrm{LCP}_\mathrm{exc} - P^\mathrm{RCP}_\mathrm{exc}$:
\begin{align}
    \label{eq:exc_CD1}
    P^\mathrm{CD}_\mathrm{exc} &= - S^2(\omega_\mathrm{p} - E^\lambda_\mathrm{exc}) 
    \nonumber \\ &\quad \times
    \int \frac{d\vec{k}}{V_\mathrm{BZ}} \mathrm{Im}[A^x_{\alpha\beta}(\vec{k}) A^y_{\beta\alpha}(\vec{k})] \left| Y^\lambda_{\alpha\beta}(\vec{k})\right|^2 \ .
\end{align}
Here, $V_\mathrm{BZ}$ is the area of the BZ.
For TMDCs, the quantum geometry in the vicinity of the K/K' valleys is determined by the top valence ($\beta$) and the bottom conduction ($\alpha$) band~\cite{fang_ab_2015}. As a consequence, the Berry connections can be related to the Berry curvature, yielding
\begin{align}
    \label{eq:exc_CD2}
    P^\mathrm{CD}_\mathrm{exc} &= - \frac12 S^2(\omega_\mathrm{p} - E^\lambda_\mathrm{exc})
    \int \frac{d\vec{k}}{V_\mathrm{BZ}} 
    \Omega_{\alpha}(\vec{k}) \left| Y^\lambda_{\alpha\beta}(\vec{k})\right|^2 \ .
\end{align}
The distinct Berry curvature texture in monolayer TMDCs (see Fig.~\ref{fig:berry_fourier}(a)) thus determines the exciton population induced by circularly polarization light, giving rise to valley polarization.  
Based on this close connection, we can track the signatures of the quantum geometry:
the dichroic exciton population and the exciton envelope function (which can be determined independently~\cite{Dong21}) directly correspond to the Berry curvature texture in the case of two relevant bands (for more bands the correspondence stays intact qualitatively ). In particular, the strongly localized nature of $Y^\lambda_{\alpha\beta}(\vec{k})$~\cite{Dong21}  effectively limits the BZ integral in Eq.~\eqref{eq:exc_CD2} to either the K or K' valley. While absolute numbers can only be extracted using accurate theory input, the positive-negative texture of the dichroic exciton population is directly proportional to the texture of the Berry curvature.

We are now ready to analyze the Fourier transform of the measured polarization-modulated photoemission intensities (Eq.~\eqref{eq:fourier_intensity}), in an energy- and momentum-resolved fashion. In particular, the $n=1$ component reflects the circular dichroism (Eq.~\eqref{eq:fourier_n1_cd}), which should directly reflect the Berry curvature texture (Eq.~\eqref{eq:exc_CD2}). Indeed, the imaginary part $\mathrm{Im}[I_{n=1}(k_x,k_y,E)]$, energy-integrated over the spectral region where the excitons peaks occur,  (Fig.~\ref{fig:berry_fourier}(d)) shows clear dichroic features at the K/K' valleys. The alternating positive-negative pattern matches exactly the behavior of the in the conduction band Berry curvature (Fig.~\ref{fig:berry_fourier}(a)).

The Fourier component $\mathrm{Im}[I_{n=1}(k_x,k_y,E)]$ obtained with our theoretical calculations (Fig.~\ref{fig:berry_fourier}(f)) is in very good agreement with the experiment. We obtain the identical positive-negative pattern which -- within the theory -- can exactly be traced back to the momentum dependence of the Berry curvature (see Eq.~\eqref{eq:exc_CD2}). The width of the peaks is governed by the exciton envelope function $Y^\lambda_{\alpha\beta}(\vec{k})$.

\subsection{Role of ultrafast scattering processes}

Apart from the bright excitons manifesting in the trARPES signal at K/K', the experimental data clearly 
feature additional excited states signals around the $\Lambda$/$\Lambda'$ valleys. Despite being clearly weaker than at the K/K' valleys, these features are characterized by the same alternating sign pattern between adjacent $\Lambda$/$\Lambda'$ valleys. The origin of the population at the $\Lambda$/$\Lambda'$ valleys is well known: its originates from K-$\Lambda$ inter-valley scattering, leading to the formation of momentum-forbidden dark excitons, with electron and hole residing at the $\Lambda$ and K valleys, respectively. Because of their momentum-indirect nature, these excitons cannot be prepared by a direct (vertical) optical transition. Understanding the origin of the $\Lambda$/$\Lambda'$ valleys $\mathrm{Im}[I_{n=1}(k_x,k_y,E)]$ texture thus requires more sophisticated modeling, including ultrafast scattering processes following photoexcitation. 

Indeed, electron-phonon and electron-electron scattering limit the lifetime of the bright excitons. Two mechanisms are dominant on tens of femtosecond time scale: (i) electrons scattering to the $\Lambda$ valleys, and (ii) electrons scattering from K to K' (or K' to K)~\cite{Maialle93, Schmidt16}. The spin polarization and the Berry curvature are locked, and adjacent K/K' valleys are characterized by opposite spin- and Berry curvature textures (see Fig.~\ref{fig:berry_fourier}(b)). These properties strongly influence the ultrafast exciton dynamics in 2D systems~\cite{Bertoni16, Madeo20, Dong21,fanciulli2023ultrafast}. To compare experiment and theory directly, we solved a quantum-master equation:
\begin{align}
    \label{eq:lindblad}
    \frac{d}{dt} \gvec{\rho}(t) &= -i [\vec{H}(t),\gvec{\rho}(t)] + \sum_n \gamma_n \vec{D}_n[\gvec{\rho}(t)] \ .
\end{align}
Here, $\gvec{\rho}(t)$ is the many-body density matrix in the space of the ground state (index $\nu=0$) and the bright ($\nu=1,2$, corresponding to $\vec{p}=0$) and dark ($\nu > 2$, corresponding to $\vec{p} \ne 0$) excitons. We can thus identify $P^\lambda_\mathrm{exc}(\vec{p},t) = \rho_{\nu\nu}(t)$ for $\nu > 0$.
The scattering operators $\vec{D}_n[\gvec{\rho}]$ ($n$ labels the scattering channels) are constructed such that they incorporate (i) K$\leftrightarrow$K' scattering (rate $\gamma_n = T^{-1}_{\mathrm{K}\rightarrow \mathrm{K'}}$), (ii) K$\rightarrow \Lambda$ scattering (rate $\gamma_n = T^{-1}_{\mathrm{K}\rightarrow \Lambda}$), and (iii) general dephasing of the off-diagonal components (rate $\gamma_n = T^{-1}_\mathrm{deph}$). The diagonal components of the time-dependent exciton Hamiltonian are given by the exciton energies $E_\nu = E^\lambda_\mathrm{exc}(\vec{p})$, while the off-diagonal elements $H_{\nu 0}(t) = -\vec{E}_\mathrm{IR}(t)\cdot \vec{M}^\lambda$ (for $\nu$ denoting the bright excitons) describe the light-matter coupling. Substituting the exciton population obtained from solving the master equation~\eqref{eq:lindblad} (averaging over the duration of the probe pulse) into the trARPES expression~\eqref{eq:trARPES_theory} yields an excellent match with the experimental exciton (polarization-averaged) intensity (Fig.~\ref{fig:berry_fourier}(c), (g)) for $ T_{\mathrm{K}\rightarrow \mathrm{K'}}=120$~fs and $T_{\mathrm{K}\rightarrow\Lambda} = 80$~fs. The only major difference is the intensity peak around the $\Gamma$ point observed in the experiments, which is attributed to LAPE~\cite{Glover96}. Similarly, the agreement between experiment and theory is improved for the $n=1$ Fourier component (Fig.~\ref{fig:berry_fourier}(d), (h)).

Strikingly, despite being significantly weaker, the dichroism encoded in the $n=1$ Fourier component from the $\Lambda$ valleys has the same sign as the dichroism at the closest K or K' valley. While the Berry curvature texture in the $\Lambda$ valleys is roughly similar to the corresponding K/K' valley, it possesses a pronounced momentum dependence (weaker for smaller parallel momenta), which is not observed in the experiments nor in the theory. Indeed, the dichroism is determined by the pump-induced population, i.e. by the interband vertical optical transitions. With LCP (RCP) polarization, the spin-polarized electrons forming the bright excitons at K (K') scatter to $\Lambda$ valleys with the same spin, while spin-flip processes have a low probability (see Fig.~\ref{fig:berry_fourier}(b))~\cite{fanciulli2023ultrafast}. Therefore, the valley selectivity of the pump-induced bright exciton population is preserved by the K$\rightarrow\Lambda$ (K'$\rightarrow\Lambda$) scattering process, due to the constrain on scattering pathways imposed by the spin texture. This "memory" effect is also present in our calculations (Fig.~\ref{fig:berry_fourier}(h)), confirming this physical mechanism.

\begin{figure}[t]
    \centering
    \includegraphics[width=\columnwidth]{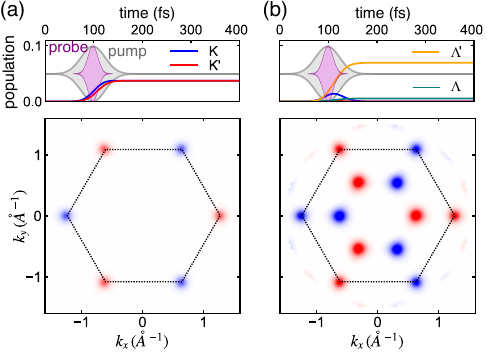}
    \caption{\textbf{Impact of ultrafast scattering on the dichroism.} (a) Time-dependent population of the exciton states upon pumping with LCP light in normal incidence, along with the envelope of the pump pulse and the probe pulse (top panel), and corresponding energy-integrated Fourier signal $\mathrm{Im}[I_{n=1}(k_x,k_y)]$ for $T_{\mathrm{K}\rightarrow \mathrm{K}^\prime} = 10$~fs, $T_{\mathrm{K}\rightarrow\Lambda} = \infty$. (b) Same as (a), but for  $T_{\mathrm{K}\rightarrow \mathrm{K}^\prime} = \infty$ and $T_{\mathrm{K}\rightarrow\Lambda} = 10$~fs. The color scale is consistent with Fig.~\ref{fig:berry_fourier}(f),(h).}
    \label{fig:scattering}
\end{figure}

In contrast, post-optical transition ultrafast intervalley scattering involving spin-flip processes would reduce the measured dichroism. In particular, K$\leftrightarrow$K' (or vice-versa) scattering would give rise to electron populations in the minority valley, thus leading to a weaker polarization modulation of the valley-resolved population. While it is very challenging to control them experimentally, our theoretical approach allows us to investigate the role of scattering processes by tuning their characteristic times $T_{\mathrm{K}\rightarrow\mathrm{K'}}$ and $T_{\mathrm{K}\rightarrow\Lambda}$ (Fig.~\ref{fig:scattering}).

We first investigate the situation where only $\mathrm{K}\rightarrow \mathrm{K}^\prime$ scattering channel is activated (i.e. $\mathrm{K}\rightarrow \Lambda$ is forbidden -- $T_{\mathrm{K}\rightarrow\Lambda} = \infty$ -- see Fig.~\ref{fig:scattering}(a)). In this case, the population of the excitons localized at K/K' approach the same value rapidly, thus reducing the dichroic signal. Note that even for scattering times as fast as  $T_{\mathrm{K}\rightarrow \mathrm{K}^\prime} = 10$~fs, which has been used for the simulation in Fig.~\ref{fig:scattering}(a), the dichroism is not fully suppressed. Ultrafast scattering processes thus blur the direct correspondence between the momentum-resolved optical transition rate and the Berry curvature. 

In Fig.~\ref{fig:scattering}(b), we investigate another extreme scenario with ultrafast K$\rightarrow \Lambda$ scattering ($T_{\mathrm{K}\rightarrow\Lambda} = 10$~fs) and forbidden $\mathrm{K}\rightarrow \mathrm{K}^\prime$ channel ($T_{\mathrm{K}\rightarrow \mathrm{K}^\prime} = \infty$). In this case, the dichroic trARPES signal from the $\Lambda$ valleys dominates. Similar to Fig.~\ref{fig:scattering}(a), the quantum geometric texture still leaves its imprint onto the dichroic $\mathrm{Im}[I_{n=1}(k_x,k_y)]$ signal, despite the rapid population transfer.

\section{Discussion}
Our joint experimental and theoretical work introduces a robust scheme to extract local quantum geometric properties of the electronic structure of materials using momentum-resolved many-body optical transition rates, here exemplified for a TMDC monolayer (WSe$_2$). Indeed, we exploit the direct relationship between chiroptical selection rules for bright excitons and their Berry curvature to design a viable measurement protocol to access its texture in reciprocal space. Using continuous pump polarization modulation in trARPES in an analogous fashion to the lock-in detection scheme, we isolate signals modulated at the helicity-swap frequency. This measurement scheme allows for extracting a pure optical circular dichroism signal, efficiently removing all contamination coming from linear pump contributions, experimental geometry, or other experimental artifacts. This Fourier analysis protocol is particularly interesting for ARPES measurements, which are performed at off-normal angles of incidence, leading to non-trivial experimental geometric effects competing with intrinsic signals of interest.  

Our theoretical model allowed us to investigate the resilience of our dichroic signal towards ultrafast scattering following optical transitions. Ultrafast  reorganization of populations in energy- and momentum-space may blur the one-to-one correspondence between momentum-resolved optical transition rate and Berry curvature. However, even in the scenario where the scattering time is shorter than the pulse duration, our calculations demonstrate that the quantum geometric texture still leaves its imprint onto the dichroic $\mathrm{Im}[I_{n=1}(k_x,k_y)]$ signal. With sub-50 fs temporal resolution routinely available in trARPES setups, this measurement scheme can be applied to a wide range of material systems.

It is also interesting to mention that a simple extension of our scheme would be compatible with the recent proposal to experimentally measure the quantum metric~\cite{von_gersdorff_measurement_2021}, i.e. the real part of the quantum geometric tensor (Berry curvature is the imaginary part of the quantum geometric tensor). A light-matter interaction-based protocol to measure the quantum metric would be highly desirable, as this momentum-resolved quantity has been predicted to be of capital importance in the emergence of a broad range of physical phenomena, e.g. anomalous Hall effect~\cite{Gao14}, orbital magnetic susceptibility~\cite{Gao15}, exciton Lamb shift~\cite{Srivastava15}, as well as superconductivity~\cite{Peotta15}.

Moreover, being intrinsically compatible with ultrafast time-resolved measurements, extending our scheme to a three-pulses trARPES approach would allow measuring ultrafast light-induced modification of local quantum geometric properties of solids undergoing dynamics. 

\section{Methods}
\subsection{Experiments}
The optical setup underlying our time- and angle-resolved photoemission spectroscopy experiments is based on a home-built optical parametric chirped-pulse amplifier (OPCPA). The OPCPA is delivering up to 30 $\mu$J/pulses (15 W, 800 nm, 30 fs) at 500 kHz repetition rate~\cite{Puppin15}. In the probe arm, the second harmonic (SHG) of the OPCPA output (400 nm) is used to drive high-order harmonic generation (HHG) by tightly focusing (15 $\mu$m FWHM) laser pulses onto a thin and dense Argon gas jet. The nonlinear interaction between the laser pulses and the Argon atoms leads to the generation of a comb of odd harmonics of the driving laser, extending up to the 11th order. A single harmonic (7th order, 21.7 eV) is isolated by reflection off a focusing multilayer XUV mirror and transmission through a 400 nm thick Sn metallic filter. A photon flux of up to 2x10$^{11}$ photons/s at the sample position is obtained (110 meV FWHM)~\cite{Puppin19}. As a pump beam, we used s-polarized near-infrared pulses (760~nm, $\hbar\omega_\mathrm{IR}=1.63$~eV, $\sim$ 45~fs full width at half maximum (FWHM) duration), to resonantly prepare bright A-excitons in ML-WSe$_2$ sample. We use a quarter-wave plate located before the pump and probe recombination chamber to control the polarization state of the pump pulse. The NIR pump and XUV probe pulses are noncollinear recombined and focused onto the sample lying in the photoemission end-station. The photoemission data are acquired using a time-of-flight momentum microscope (METIS1000, SPECS GmbH), allowing to detect each photoelectron as a single event, as a function of NIR quarter-waveplate angle ($\theta$). The resulting 4D photoemission intensity data have the coordinates $I(k_x, k_y, E, \theta)$.

Concerning the preparation of the atomically thin TMDC sample, first, thin hBN flakes are mechanically exfoliated on polydimethylsiloxane (PDMS) and transferred onto a 0.5 wt$\%$ Nb-doped rutile TiO$_2$ (100) substrate. Subsequently, monolayer WSe$_2$ is exfoliated from bulk crystals (HQ graphene) on PDMS and stamped on top of the previously transferred hBN flake. The sample is then annealed in a high vacuum at 180$^{\circ}$C for at least 2h at each step. The hBN serves as an atomically smooth buffer layer to prevent the corrugation of substrate surface roughness~\cite{man2016protecting}, and the slightly conductive substrate TiO$_2$ reduces the space charging effect from trARPES measurements~\cite{ulstrup2016spatially}. 

\subsection{First-principle calculations}

We performed density-functional theory (DFT) calculations with the full-electron code \textsc{FLEUR}~\cite{fleurCode} within the Perdew-Burke-Ernzerhof (PBE) approximation~\cite{perdew_generalized_1996} to the exchange-correlation functional and subsequently constructed projective Wannier functions $\phi_j(\vec{r})$ using the \textsc{Wannier90} code~\cite{pizzi_wannier90_2020}. We included the W$-d$ and the Se-$p$ orbitals. As the next step we performed a one-shot $G^0W^0$ calculation \cite{hedin_new_1965} to obtain the self-energy $\Sigma_{\alpha}(\vec{k},\omega)$, from which the quasiparticle energies $\varepsilon_\alpha(\vec{k})$ are computed. The resulting quasiparticle Hamiltonian is expressed in the Wannier basis, yielding an 11-orbital model reproducing the $G^0W^0$ bands with high accuracy. 

As the next step, we performed constrained random-phase approximation (cRPA) calculations~\cite{aryasetiawan_frequency-dependent_2004} to obtain the Coulomb matrix elements in the Wannier basis using the SPEX code~\cite{friedrich_efficient_2010}. Due to reduction to the bands spanned by the Wannier functions, the Coulomb interaction attains a frequency dependence. However, as the energy scale of the screening effects is much bigger than the band gap, we approximate the interaction as static ($\omega=0$). Furthermore, we only keep the density-density matrix elements due to the localized nature of the Wannier functions. Thus we obtain the interaction Hamiltonian
\begin{align}
    \label{eq:ham_int}
    \hat{H}_\mathrm{int} = \frac12 \sum_{\vec{R},\vec{R}^\prime} \sum_{jj^\prime} U_{jj^\prime}(\vec{R} - \vec{R}^\prime) \hat{n}_{\vec{R}j} \hat{n}_{\vec{R}^\prime j^\prime } \ ,
\end{align}
where $\hat{n}_{\vec{R}j} = \hat{c}^\dagger_{\vec{R}j} \hat{c}_{\vec{R}j}$ is the density operator for the lattice site $\vec{R}$ and orbital $j$. The Coulomb interactions $U_{jj^\prime}(\vec{R} - \vec{R}^\prime)$ are presented in the supplemental materials, along with full details of the calculations.

\subsection{Wannier model}

With the $G^0 W^0$-Wannier Hamiltonian and the Coulomb interactions, we have a flexible and accurate model for the electronic structure, including excitons. To obtain the exciton envelope function, we solved the Wannier equation~\cite{haug_quantum_1990,berghauser_analytical_2014} for a selected pair of valence ($\beta$) and conduction ($\alpha$) bands:
\begin{align}
    \Big[\varepsilon_\alpha(\vec{k} + \vec{p}) - &\varepsilon_\beta(\vec{k}) - E^\lambda_\mathrm{exc}(\vec{p})\Big]  Y^{\lambda}_{\alpha\beta}(\vec{p},\vec{k})
    \nonumber \\ 
   & \quad - \sum_{\vec{q}} W_{\alpha\beta}(\vec{k} + \vec{p},\vec{k} + \vec{q},\vec{q}) Y^{\lambda}_{\alpha\beta}(\vec{p},\vec{q}) = 0 \ .
\end{align}
The effective interaction $W_{\alpha\beta}(\vec{k},\vec{k}^\prime,\vec{p})$ is the inter-band screened interaction. As the precise dielectric environment of the substrate is hard to characterize, we employed the effective continuum model from refs.~\cite{rosner_wannier_2015,steinke_coulomb-engineered_2020}. The model dielectric function $\epsilon(\vec{q})$ is parameterized by the dielectric constant at $\omega\rightarrow \infty$, $\epsilon_\infty$, the substrate dielectric function $\epsilon_\mathrm{sub}$, and the effective thickness of the WSe$_2$ layer $d_\mathrm{eff}$. We fixed $d_\mathrm{eff} = 6.48$~\AA\,  as in ref.~\cite{steinke_coulomb-engineered_2020} while adjusting $\epsilon_\infty$ and $\epsilon_\mathrm{sub}$ to match the exciton binding energies observed in the experiments. The resulting absorption spectrum (see supplemental materials) is in good agreement with first-principle calculations for WSe$_2$ on hBN substrate.

Once the exciton envelope functions $Y^{\lambda}_{\alpha\beta}(\vec{p}\vec{k})$ (we take the lowest states $\lambda$ only) have been determined, optical matrix elements are computed as $\vec{M}^\lambda_\mathrm{exc} = \delta_{\vec{p},0}\sum_{\vec{k}\alpha\beta} Y^{\lambda}_{\alpha\beta}(\vec{p},\vec{k})\vec{A}_{\alpha\beta}(\vec{k})$. The Berry connections $\vec{A}_{\alpha\beta}(\vec{k})$ are calculated from the Wannier Hamiltonian as in ref.~\cite{yates_spectral_2007}. 

\subsection{Time-dependent dynamics}

To simulate the population dynamics we derived the quantum-master equation~\eqref{eq:lindblad} from the Lindblad formalism. Thus, the scattering operators are constructed as
\begin{align}
    \label{eq:scatt_oper}
    \vec{D}_n[\gvec{\rho}] = \vec{L}_n \gvec{\rho} \vec{L}^\dagger_n - \frac{1}{2}\left\{\vec{L}^\dagger_n \vec{L}_n , \gvec{\rho} \right\} \ ,
\end{align}
where $\{ , \}$ denotes the anti-commutator. The Lindblad operators are constructed as projectors as follows: (i) $\vec{L}_n = |\Psi^\mathrm{exc}_{0\lambda}\rangle \langle \Psi^\mathrm{exc}_{\vec{p}\lambda^\prime}| $ for the scattering process from K/K' (corresponding to $\nu=1,2$) to the dark exciton states with corresponding momentum $\vec{p}$ ($\nu^\prime > 2$), (ii) $\vec{L}_n = |\Psi^\mathrm{exc}_{0\lambda}\rangle \langle \Psi^\mathrm{exc}_{0\lambda^\prime}| $ for the K$\leftrightarrow$K' process with $\nu=1,2$, $\nu^\prime = 2,1$, and (ii) $\vec{L}_n = |\Psi_0\rangle \langle \Psi_0| + \sum_{\vec{p}\lambda} |\Psi^\mathrm{exc}_{\vec{p}\lambda}\rangle \langle \Psi^\mathrm{exc}_{\vec{p}\lambda}| $ to capture the dephasing of off-diagonal components of the density matrix. We fix $T_\mathrm{deph}=40$~fs for all calculations.

Inserting the scattering operators~\eqref{eq:scatt_oper}, the optical transition matrix elements $\vec{M}^{\lambda}$, and the pump pulse with parameters consistent with the experiments into the master equation~\eqref{eq:lindblad} yields the time-dependent density matrix $\rho_{\nu\nu}(t)$, from which the trARPES spectra presented in the text are computed. 

\section{Acknowledgments}

\textbf{Funding:} This work was funded by the Max Planck Society, the European Research Council (ERC) under the European Union’s Horizon 2020 research and innovation program (Grant No. ERC452 2015-CoG-682843), H2020-FETOPEN-2018-2019-2020-01 (OPTOLogic—grant agreement No. 899794)), the German Research Foundation (DFG) within the Emmy Noether program (Grant No. RE 3977/1), the priority program SPP2244 (project 443366970 and 443405595), the SFB/TRR 227 ”Ultrafast Spin Dynamics” (projects A09 and B07), and the Würzburg-Dresden Cluster of Excellence on Complexity and Topology in Quantum Matter (ct.qmat) (EXC 2147, Project-ID 390858490). 
This research was also supported by the NCCR MARVEL, a National Centre of Competence in Research, funded by the Swiss National Science Foundation (grant number 205602).
K.W. and T.T. acknowledge support from the JSPS KAKENHI (Grant Numbers 20H00354, 21H05233, and 23H02052) and World Premier International Research Center Initiative (WPI), MEXT, Japan. T.P. acknowledges funding from the Alexander von Humboldt Foundation. M.S. acknowledges support from SNSF Ambizione Grant No. PZ00P2 193527. P.W. acknowledges support from ERC Consolidator Grant No. 724103. S.B. acknowledges support from ERC Starting Grant ERC-2022-STG No. 101076639. 

\textbf{Author contributions:} S.B. and S.D. contributed equally to this work. S.B. and M.S. conceived the idea. S.D. performed the experiments. S.B. and M.S. analyzed the experimental data. R.E., L.R., and M.W. were responsible for developing the experimental infrastructures. S.B., S.D., and T.P. participated in maintaining and running the experimental apparatus. J.D.Z and A.C. prepared the ML sample with the hBN substrate provided by T.T. and K.W.. M.S. developed the theory with inputs from V.C. and P.W.. S.B. and M.S. wrote the first draft of the manuscript. 

\textbf{Competing interests:} The authors declare that they have no competing interests.

\textbf{Data and materials availability:}  Raw trARPES data along with \textsc{Python} scripts for post-processing and visualization, as well as the \textsc{Python} scripts producing the figures in the main text can be found at Materials Cloud Archive 2023.128 (2023), doi: 10.24435/materialscloud:zq-tj. The custom computer code to solve the Wannier equation is based on the \textsc{dynamics-w90} code and will be made available upon reasonable request.


\providecommand{\noopsort}[1]{}\providecommand{\singleletter}[1]{#1}%

\end{document}


\maketitle

\renewcommand{\thefigure}{S\arabic{figure}}
\renewcommand{\thetable}{S\arabic{table}}

\section*{Supplementary Note 1: Computational details for the first principles calculations}

We start from a density functional theory (DFT) calculation of monolayer WSe$_2$ within the generalized gradient approximation (GGA)~\cite{perdew_generalized_1996} using the full-potential all-electron code \textsc{Fleur}~\cite{fleurCode}. We use a $32\times32\times1$ ${\bf k}$-grid for the calculations, and to avoid interactions with the periodic image along the $z$ (nonperiodic) direction we use a 23~\AA~vacuum distance. 
The primitive cell contains a single formula unit, and the atomic positions and lattice parameters for the hexagonal lattice are given in Tab.~\ref{tab:lattice_parameters}.

\begin{table*}[h]
\caption{Lattice parameters in \AA~and atomic positions in Cartesian coordinates for the relaxed structure of hexagonal monolayer WSe$_2$.
\label{tab:lattice_parameters}
}
\setlength{\tabcolsep}{15pt} 
\renewcommand{\arraystretch}{0.81} 
\centering
\begin{tabular}{|c|c||c|c|c||c|c|c|}
\multicolumn{4}{c}{\vspace*{-0.5cm}} \\
\hline
\multicolumn{4}{c}{$a$=3.32~\AA, $c=8a$}    \\
\hline
& $x$ [\AA] & $y$ [\AA] & $z$ [\AA]\\
\hline
\hline
W & 0 & 0 & 0  \\
Se & 1.660 & -0.958 & -1.675  \\
Se & 1.660 & -0.958 & 1.675  \\
\hline
\end{tabular}
\end{table*}

\begin{figure}[h]
\centering
\includegraphics[width=0.96\columnwidth]{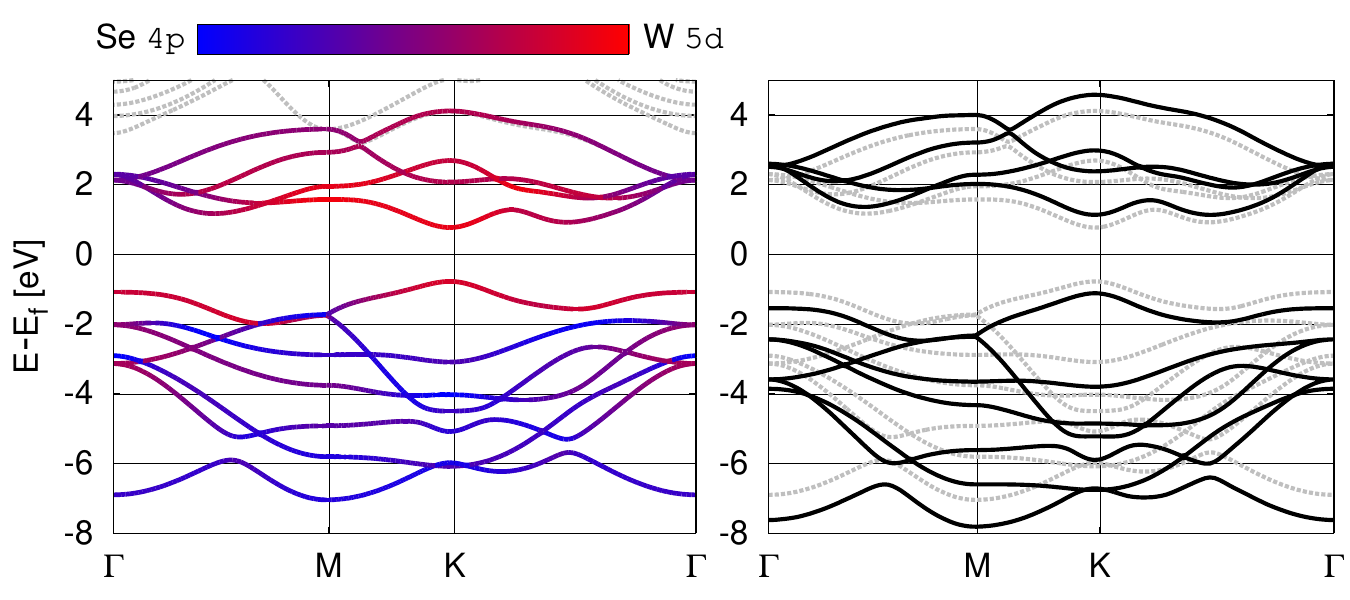}
\caption{\textbf{Band structure of monolayer WSe$_2$.} Left: The model band structure (solid lines) plotted on top of the DFT band structure (gray dashed lines). The color bar shows the orbital-like character of the model bands. Right: The $G^0W^0$ quasiparticle band structure (black lines) and disentangled model bands (gray dashed lines). The Fermi energy is placed in the middle of the gap.
\label{fig:BS}}
\end{figure}

We construct a low-energy model subspace using projected Wannier functions from the Wannier90 library~\cite{pizzi_wannier90_2020} as basis functions. We include W $5d$- and Se $4p$-like orbitals for the three sites in the primitive cell, defining a model spanned by eleven orbitals in total. The resulting disentangled band structure agrees very well with the DFT calculation, as shown in Fig.~\ref{fig:BS}. The model should therefore provide a good starting point for our many-body calculations.

To define a quasiparticle $GW$ Hamiltonian, we first perform a one-shot $G^0W^0$ calculation \cite{hedin_new_1965} to compute the self-energy $\Sigma_{\alpha}(\vec{k},\omega)$. The quasiparticle energies are then obtained by solving the quasiparticle 
equation~\cite{hybertsen_electron_1986}
\begin{equation}
\varepsilon^\textrm{QP}_{\alpha}(\vec{k}) = \varepsilon_{\alpha}(\vec{k}) - V^\textrm{xc}_{\alpha}(\vec{k}) + \mathrm{Re}\left[\Sigma_{\alpha}(\vec{k},\en^\textrm{QP}_{\alpha}(\vec{k}))\right] \ ,
\end{equation}
where $\varepsilon_{\alpha(\vec{k})}$ and $V^\textrm{xc}_{\alpha}(\vec{k})$ are the Kohn-Sham eigenvalues and the exchange-correlation potential from the DFT calculation, respectively. The resulting $G^0W^0$ quasiparticle band structure is shown in Fig.~\ref{fig:BS},  where we in particular note the usual widening of the band gap.

Within the constrained random-phase approximation (cRPA) \cite{aryasetiawan_frequency-dependent_2004}  the effective bare interaction $U$ for a low-energy subspace is calculated as
\begin{equation}\label{eq:UcRPA_Pr}
U(\omega)=[1-v\Pi_r(\omega)]^{-1} v.
\end{equation}
Here $v$ is the bare Coulomb interaction and the screening from within the $dp$-model, $\Pi_d$, has been removed from the polarization function,
$\Pi_r=\Pi-\Pi_d$. Through this downfolding of the higher-energy states, the effective interaction for the low-energy model is defined as
\begin{align}\label{eq:U_wannier}
U_{j_1j_2j_3j_4}(\omega,{\bf R}) &= \int \int \mathrm{d}{\bf r}\mathrm{d}{\bf r}' \phi_{j_1{\bf 0}}^*({\bf r})\phi_{j_2{\bf 0}}({\bf r})U({\bf r},{\bf r}',\omega)\phi^*_{j_3{\bf R}}({\bf r}')\phi_{j_4{\bf R}}({\bf r}') \nonumber \\
&= \frac{1}{N}\sum_{\bf{q}} e^{-i \bf{q}\bf{R}} U_{j_1j_2j_3j_4}(\omega,{\bf q})
\end{align} 
attains a frequency dependence, in addition to being a nonlocal quantity. The basis functions $\varphi_{j{\bf 0}}({\bf r})$ are here taken to be the localized Wannier functions defining the model subspace.
In this study, we have chosen to retain the full nonlocal nature,
while limiting ourselves to only treating the static ($\omega=0$) density-density interaction terms: $U_{j j^\prime}(\vec{R}) \equiv U_{jj,j^\prime j^\prime}(\omega=0,{\bf R})$.

\begin{figure}[t]
\centering
\includegraphics[width=0.86\columnwidth]{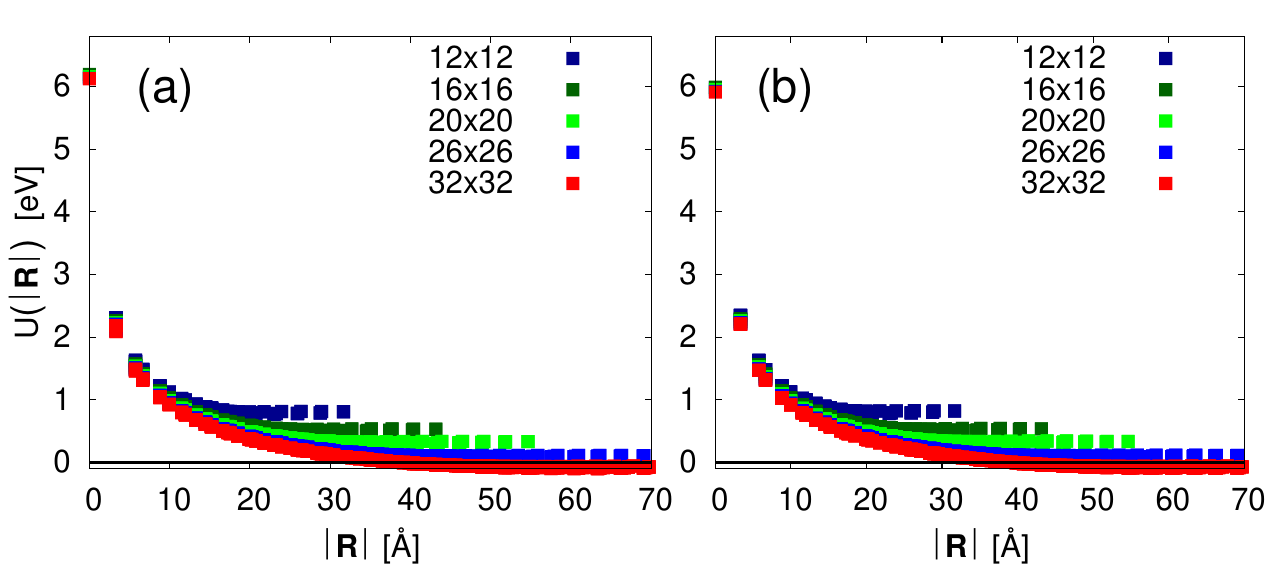}
\caption{\textbf{Distance dependence of the effective interactions.} The static nonlocal cRPA interaction $U(\omega=0,|{\bf R}|)$ calculated with varying ${\bf k}$-grids (see the labels) for (a) one of the W $5d$-like orbitals and (b) one of the Se $4p$ -like orbitals.
\label{fig:interactions}}
\end{figure}

Since the metallic screening coming from the low-energy region is removed, the calculated effective bare interaction typically displays a long-ranged Coulomb-like tail. It is therefore important that both the local (${\bf R}=0$) and long-range parts are well-converged with respect to the momentum grid, in particular as we for a 2D system can expect the remaining screening to be less efficient. In Fig.~\ref{fig:interactions} we show the nonlocal interaction $U(|{\bf R}|)$ for one W $5d$- and one Se $4p$-like orbital calculated using increasingly dense momentum grids. We can note that while the local and short-range interaction terms converge rapidly, the long-range tail is more sensitive to the sampling.
We have further checked that increasing the vacuum distance to 30~\AA~only produces a small ($\sim0.1$ eV) change in the local quantities for the $20\times20\times1$ {\bf k}-grid, which is within our ``errorbar" for these calculations.

The cRPA and $G^0W^0$ calculations are performed using the SPEX code~\cite{friedrich_efficient_2010}. The quantities used for the results presented in the main text have been calculated using the dense $32\times32\times1$ ${\bf k}$-grid, and we include DFT bands up to $\sim40$ eV when computing both the polarization function and self-energy.

\section*{Supplementary Note 2: Wannier equation and exciton properties}

With the compact representation of the Coulomb interaction in the Wannier basis, $U_{j j^\prime}(\vec{R})$, we can express any inter-band interaction. For excitons, the relevant contribution is the interband interaction

\begin{align}
\label{eq:ham_int_bare}
 \hat{H}_\mathrm{int} &= \sum_{\alpha \beta } \sum_{\textbf{k}_1 \textbf{k}_2 \textbf{q}} \left[\sum_{j j^\prime} U_{j j^\prime}(\vec{q})  C_{j \alpha}^{*}(\vec{k}_1) C_{j^\prime \beta}^{*}(\vec{k}_2)  C_{j^\prime\alpha}(\vec{k}_2+\vec{q}) C_{j \beta}(\vec{k}_1 - \vec{q}) \right] \hat{c}_{\alpha \textbf{k}_1}^\dagger  \hat{c}_{\beta \textbf{k}_2}^\dagger \hat{c}_{\beta \textbf{k}_2 +\textbf{q}}  \hat{c}_{\alpha \textbf{k}_1-\textbf{q}} \nonumber
 \\ &\equiv \sum_{\alpha \beta } \sum_{\textbf{k}_1 \textbf{k}_2 \textbf{q}} V_{\alpha\beta}(\vec{k}_1,\vec{k}_2,\vec{q}) \hat{c}_{\alpha \textbf{k}_1}^\dagger  \hat{c}_{\beta \textbf{k}_2}^\dagger \hat{c}_{\beta \textbf{k}_2 +\textbf{q}}  \hat{c}_{\alpha \textbf{k}_1-\textbf{q}} \ .
\end{align}
Here, $\alpha$ ($\beta$) label the conduction (valence) bands. We have approximated the interaction as diagonal in band space.  

Exciton properties are obtained from solving the Wannier equation with the screened interaction, while Eq.~\eqref{eq:ham_int_bare} describes the bare interaction in the Wannier space. To build in the screening, we employ the continuum model from refs.~\cite{rosner_wannier_2015,steinke_coulomb-engineered_2020}, which allows us to incorporate the screening of the hBN substrate in a phenomenological way. The model dielectric function $\epsilon(\vec{q})$ (we only the consider static screening) is parameterized by the dielectric function for $\omega\rightarrow \infty$ ($\epsilon_\infty$), the substrate dielectric function ($\epsilon_\mathrm{sub}$), and the effective width of the WSe$_2$ layer ($d_\mathrm{eff}$). As in ref.~\cite{steinke_coulomb-engineered_2020} we fix $d_\mathrm{eff} = 6.48$~a.u., while we chose $\epsilon_\infty = 2$ and $\epsilon_\mathrm{sub} = 3.375$. We thus obtain the statically screened interaction
\begin{align}
    \label{eq:screenedW}
    W_{\alpha\beta}(\vec{k}_1,\vec{k}_2,\vec{q}) = \frac{V_{\alpha\beta}(\vec{k}_1,\vec{k}_2,\vec{q})}{\epsilon(\vec{q})} \ .
\end{align}

The exciton states are treated within the particle-hole expansion
\begin{align}
    \label{eq:psi_exc}
    |\Psi^\mathrm{exc}_{\vec{p}\lambda}\rangle = Y^{\lambda}_{\alpha\beta}(\vec{p},\vec{k}) \hat{c}^\dagger_{\vec{k}+\vec{p}\alpha}\hat{c}_{\vec{k}\beta} | \Psi_0\rangle \ ,
\end{align}
where $|\Psi_0\rangle$ is the quasi-particle ground state. Inserting the exciton state into the Schr\"odinger equation with the interaction Hamiltonian~\eqref{eq:ham_int_bare}, replacing $V_{\alpha\beta}(\vec{k}_1,\vec{k}_2,\vec{q}) \rightarrow W_{\alpha\beta}(\vec{k}_1,\vec{k}_2,\vec{q})$, then yields the Wannier equation for the exciton envelope function
\begin{align}
    \label{eq:bse}
    \sum_{\vec{q}} \mathcal{H}^{\vec{p}\lambda}_{\alpha\beta}(\vec{k},\vec{q}) Y^{\lambda}_{\alpha\beta}(\vec{p},\vec{q}) = E_\lambda(\vec{p}) Y^{\lambda}_{\alpha\beta}(\vec{p},\vec{k}) \ .
\end{align}
Here, the effective two-particle Hamiltonian is given by
\begin{align}
\label{eq:ham_eff}
\mathcal{H}^{\vec{p}\lambda}_{\alpha\beta}(\vec{k},\vec{q}) =\left[\en^\mathrm{QP}_\alpha(\vec{k} + \vec{p}) - \en^\mathrm{QP}_\beta(\vec{k}))\right] \delta_{\vec{k}\vec{q}} - \frac{1}{N}W_{\alpha\beta}(\vec{k} + \vec{q}, \vec{k} + \vec{p},\vec{q})  \ .
\end{align}
Solving Eq.~\eqref{eq:bse} then yields the exciton envelope function for each eigenstate $\lambda$. 

\begin{figure}[t]
    \centering
    \includegraphics[width=\columnwidth]{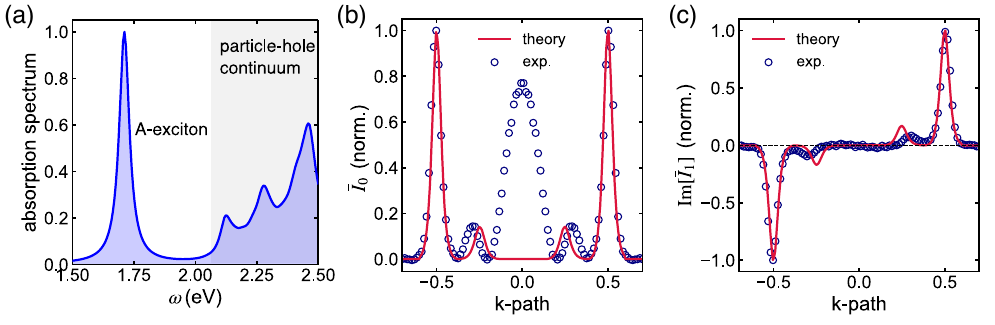}
    \caption{\textbf{Excitonic properties.} (a) Optical absorption spectrum $A(\omega)$ obtained from Eq.~\eqref{eq:absorption}. We averaged over the light polarization $\vec{e}$. (b) Energy-integrated and polarization-averaged intensity $\bar{I}_0(\vec{k})$ along the $k_x$ direction, extracted from experiments (circles) and theory (solid line). (c) Similar to (b), the imaginary component of the $n=1$ Fourier component of the energy-integrated intensity. }
    \label{fig:exciton}
\end{figure}

To benchmark our method and justify the choice of screening parameters, we have also computed the absorption spectrum 
\begin{align}
    \label{eq:absorption}
    A(\omega) = \frac{1}{\pi}\mathrm{Im} \langle \Psi_0 | \hat{D}^\dagger_{\vec{e}} \frac{1}{\omega - \hat{H} - i\eta} \hat{D}_{\vec{e}}|\Psi_0 \rangle \ , 
\end{align}
where
\begin{align}
    \hat{D}_{\vec{e}} = \sum_{\vec{k}}\sum_{\alpha\beta} \vec{e}\cdot \vec{A}_{\alpha\beta}(\vec{k}) \hat{c}^\dagger_{\vec{k}\alpha}\hat{c}_{\vec{k}\beta} 
\end{align}
denotes the dipole operator comprised of the Berry connections $\vec{A}_{\alpha\beta}(\vec{k})$ and the light polarization $\vec{e}$. Fig.~\ref{fig:exciton}(a) shows the absorption spectrum, featuring the prominent A-exciton peak and the particle-hole continuum. The exciton binding energy is in excellent agreement with the experiments, while the spectrum in Fig.~\ref{fig:exciton}(a) agrees very well with first-principle calculations for WSe$_2$ on a thick hBN substrate.

\section*{Supplementary Note 3: Fitting of the scattering times}

From the exciton envelope function we computed the dipole transition matrix elements $M^\lambda = \langle \Psi^\mathrm{exc}_{0\lambda} | \hat{D}_\vec{e} | \Psi_0\rangle $ for a specific light polarization. Combining with the exciton energies $E^\lambda(\vec{p})$, we have all ingredients to build the effective Hamiltonian in exciton space, as discussed in the methods section in the main text. To further ensure the accuracy of our approach to obtaining exciton properties and to determine the scattering times $T_{\mathrm{K}\rightarrow\Sigma}$ and $T_{\mathrm{K}\rightarrow \mathrm{K'}}$, we solved the quantum master equation (Eq. (9) in the main text) for various parameters. We adopted the pump and the probe pulse as in the experiments and also modulated the pump polarization $\vec{e}$ accordingly. For each value of  $T_{\mathrm{K}\rightarrow \mathrm{K'}}$ and $T_{\mathrm{K}\rightarrow\Sigma}$ we computed the photoemission intensity (Eq.~(4) in the main text, summing over all exciton states). 

The thus simulated trARPES intensity is compared directly to experiments. To reduce the effect of the slight misalignment of the sample, we averaged the intensity over the three different paths along the K--$\Gamma$--K' directions. Experiment and theory (performing the same Fourier analysis) are directly compared to each other in Fig.~\ref{fig:exciton}(b),(c). We note the excellent agreement of theory and experiment for the A-exciton peak. The scattering times have then been optimized to reproduce the relative weight of the bright excitons and the indirect excitons (peaks at $\Sigma$) as well as possible.



%

%